Correspondence and requests for materials should be addressed to P. T (tongpeng@issp.ac.cn) or Y. P. S (ypsun@issp.ac.cn)

# Unusual ferromagnetic critical behavior owing to short-range antiferromagnetic correlations in antiperovskite Cu$_{1-x}$NMn$_{3+x}$ (0.1≤$x$≤0.4)

Jianchao Lin[1], Peng Tong[1,*], Dapeng Cui[1], Cheng Yang[1], Jie Yang[1], Shuai Lin[1], Bosen Wang[1], Wei Tong[2], Lei Zhang[2], Youming Zou[2] & Yuping Sun[1,2,3,*]

[1]Key Laboratory of Materials Physics, Institute of Solid State Physics, Chinese Academy of Sciences, Hefei 230031, China, [2]High Magnetic Field Laboratory, Chinese Academy of Sciences, Hefei 230031, China, [3]Collaborative Innovation Center of Advanced Microstructures, Nanjing University, Nanjing 210093, China.

**For ferromagnets, varying from simple metals to strongly correlated oxides，the critical behaviors near the Curie temperature ($T_C$) can be grouped into several universal classes. In this paper, we report an unusual critical behavior in manganese nitrides Cu$_{1-x}$NMn$_{3+x}$ (0.1 ≤ $x$ ≤ 0.4). Although the critical behavior below $T_C$ can be well described by mean field (MF) theory, robust**

**critical fluctuations beyond the expectations of any universal classes are observed above $T_C$ in $x$=0.1. The critical fluctuations become weaker when $x$ increases, and the MF-like critical behavior is finally restored at $x$=0.4. In addition, the paramagnetic susceptibility of all the samples deviates from the Curie-Weiss (CW) law just above $T_C$. This deviation is gradually smeared as $x$ increases. The short-range antiferromagnetic ordering above $T_C$ revealed by our electron spin resonance measurement explains both the unusual critical behavior and the breakdown of the CW law.**

Phase transition is a core concept of current condensed matter physics. The analysis of the critical behavior provides significant information about the thermodynamic observables near the transition[1]. As a classic example to address the critical behavior, the continuous paramagnetic (PM) to ferromagnetic (FM) transition has been extensively studied. In the vicinity of a second-order FM transition, the universal scaling laws apply to the spontaneous magnetization ($M_s$) and the initial magnetic susceptibility ($\chi_0$) due to the divergence of the correlation length. Namely, $M_s$ just below the Curie temperature ($T_C$) is described by the relation, $M_s \sim (1-T/T_C)^\beta$, and $\chi_0$ just above $T_C$ follows the relation[2], $1/\chi_0 \sim (T/T_C-1)^\gamma$. Accordingly, the critical isothermal $M(H)$ at $T_C$ is described by $M(H) \sim H^{1/\delta}$. The three critical exponents, $\beta$, $\gamma$ and $\delta$, fulfill the Widom scaling relation[3], $\delta = 1 + \gamma/\beta$. The critical exponents, obtained in a wide variety of FM material systems, can be grouped into a few well-known universal classes[4], such as the mean field (MF) model ($\beta$=0.5, $\gamma$=1.0 and $\delta$=3.0), 3D Heisenberg (3DH) model ($\beta$=0.365, $\gamma$=1.386 and $\delta$=4.80), and 3D Ising (3DI) model ($\beta$ = 0.325, $\gamma$ = 1.241 and $\delta$ = 4.82). Unusual critical behaviors, characterized by critical exponents out of the expected ranges of the universal classes, challenge the scaling theory of the critical phenomena and thus have attracted considerable attention[5-11].

When the temperature is well above $T_C$, the magnetic susceptibility $\chi(T)$ of a ferromagnet can be well illustrated by the Curie-Weiss (CW) law; namely, $1/\chi(T)$ is linearly dependent on temperature[12]. However, the breakdown of the CW law has been widely observed, as evidenced by either an upturn or a downward deviation from the linear temperature dependence of $1/\chi(T)$ upon cooling. The first case, a downward deviation on $1/\chi(T)$, has frequently been observed in many materials, such as perovskite manganites[11,13,14]. This situation is usually attributed to the Griffiths phase[15], where short-range (SR) FM clusters[16] occur, grow upon cooling, and finally become long-range (LR) ordered below $T_C$. By contrast, there are relatively few studies on the upturn deviation on $1/\chi(T)$ above $T_C$. Thus far, for this type of deviation, no well-accepted explanations of the driving force have been presented, although a few mechanisms, such as FM clusters[17,18], strong exchange interaction between long-wavelength spin fluctuations[19], spin lattice coupling[20], and antiferromagnetic (AFM) correlations[21] have been proposed. The unusual critical behaviors and the deviations of the magnetic susceptibility from the CW law have attracted much research interest separately, but it is unknown whether and how the two issues are related.

Manganese nitride $CuNMn_3$ is a prototype antiperovskite compound (see Supplementary Figure S1(a) online for the antiperovskite structure). This compound and related chemically doped compounds have received considerable attention because of their various functionalities[22,23], such as negative or zero thermal expansion[24-28], nearly zero temperature coefficient of resistance[29,30,31], large magnetocaloric effect[32], giant magnetostriction[33], and FM shape memory effect[34]. Upon cooling, a weakly first-order FM transition was observed at $T_C \sim$ 143 K in CuNMn$_3$, accompanied by a cubic-tetragonal structural transition[31]. Our recent study found that, by substituting Cu sites with Mn, the FM transition and structural transformation are decoupled in Cu$_{1-x}$NMn$_{3+x}$[35]. With increasing $x$, the structural transition is gradually suppressed and disappears for $x > 0.4$. In addition, $T_C$ increases with increasing $x$. Here, we report the critical behavior for Cu$_{1-x}$NMn$_{3+x}$ with $0.1 \leq x \leq 0.4$. An upturn deviation against the CW law was observed in the $1/\chi(T)$ curves above $T_C$, which is gradually smeared as $x$ increases. For low values of $x$, the critical behavior below $T_C$ follows the MF model well, whereas enhanced critical fluctuations beyond the theoretical prediction of the 3DH model were observed above $T_C$. The critical fluctuations become weakened as $x$ increases, and consequently, for $x = 0.4$, the critical behavior both below and above $T_C$ can be described by the MF model. Furthermore, our electron spin resonance (ESR) results indicate the existence of SR AFM ordering above $T_C$, which provides the physical background for the unusual critical behavior and the deviation from the CW law observed in the $1/\chi(T)$ curves.

**Results**

**Inverse magnetic susceptibility: deviation from the Curie-Weiss behavior.** Figure 1(a) shows $1/\chi(T)$ measured at a magnetic field of 0.1 kOe under the zero-field-cooled mode for Cu$_{1-x}$NMn$_{3+x}$ with $x$=0.1, 0.2, 0.3 and 0.4. The CW law, namely the linear dependence on temperature, holds at temperatures well above $T_C$. As the temperature decreases, the CW behavior breaks down because $1/\chi(T)$ exhibits an upward departure from the linear dependence. The onset temperature of the deviation is denoted as $T^*$.

The gap between $T_C$ and $T^*$ decreases as $x$ increases, that is, the upward deviation in $x$=0.4 is less pronounced than in $x$=0.1. We note that for the parent sample CuNMn$_3$, no clear deviation from the CW law can be observed in the $1/\chi$(T) curve[31]. Figure 1(b) shows a comparison of $1/\chi$(T)s for $x$=0.2 measured at 0.1 kOe, 1 kOe, 3 kOe and 10 kOe. It is apparent that all the $1/\chi$(T) curves exhibit an upward departure from the CW behavior. Moreover, $T^*$ appears to be insensitive to the applied field. This behavior contradicts the well-known Griffiths phase, in which a downward deviation in $1/\chi$(T) is usually observed at low magnetic field and disappears at high magnetic field because of the enhanced background PM signal and/or the saturation of the FM components[14].

**Critical behavior around $T_C$.** To better understand the magnetic behavior around $T_C$, critical behavior analysis was performed for Cu$_{1-x}$NMn$_{3+x}$ with $x$=0.1, 0.3 and 0.4. The isothermal magnetization $M$(H) was measured in the vicinity of $T_C$, and Arrott plots ($M^2$-$H/M$) were derived (see Supplementary Figure S2 online). The high-field isotherms of the Arrott plots were fitted with a polynomial function and then extrapolated to the $H/M$ = 0 and $M^2$ = 0 axes to obtain $M_s$ and $\chi_0^{-1}$, respectively[6]. By fitting $M_s$(T) with the relation $M_s \sim (1-T/T_C)^\beta$ and $\chi_0^{-1}$(T) with the relation $1/\chi_0 \sim (T/T_C-1)^\gamma$, we obtained the values of $\beta$ and $\gamma$, respectively. Modified Arrott plots were then obtained as $M^{1/\beta}$ versus $(H/M)^{1/\gamma}$, which were fitted again to obtain new values of $\beta$ and $\gamma$. Then, the new critical exponents were used to make modified Arrott plots again. The above procedure was continued until the critical exponents converged to stable values. The final values of $M_s$ and $\chi_0^{-1}$ are plotted in Figure 2(a) as a function of reduced temperature ($T$-$T_C$). The obtained exponent $\gamma$ decreases from ~ 1.63 for $x$=0.1, clearly larger than the value suggested by the 3DH model (1.386) or by the 3DI model (1.24), to a value of ~1.15 for $x$=0.4, which is similar to the MF magnitude of unity (1)[5]. Nevertheless, the $\beta$ values (0.537, 0.538 and 0.488 for $x$=0.1, 0.3 and 0.4, respectively) are very close to the value (0.5) predicted by the MF theory. The large $\gamma$ values for $x$=0.1 and 0.3 are likely unrelated to the 3DH or 3DI model because the $\beta$ values do not match these models. The critical exponents can also be obtained by using a Kouvel-Fisher (KF) plot[36], in which $M_S(dM_S/dT)^{-1}$ vs. $T$ and $\chi_0^{-1}(d\chi_0^{-1}/dT)^{-1}$ vs. $T$ generate straight lines with slopes $1/\beta$ and $1/\gamma$, respectively. The KF plots for $x$=0.1, 0.3 and 0.4 are displayed in Figure 2(b). The estimated values for $\beta$ and $\gamma$ are consistent with those derived from the modified Arrott plots shown in Figure 2(a), indicating that the results of the critical behavior analysis are reliable.

The critical exponent $\delta$ can be either derived from the critical isothermal magnetization curve using the relation $M(H) \sim H^{1/\delta}$ or estimated by the Widom scaling relation given the values of $\beta$ and $\gamma$. For each composition, $M(H)$ at the temperature closest to $T_C$ is plotted in Figure 2(c) in the form of $\log(H)$-$\log(M)$. Based on the linear fitting to the high-field data, the $\delta$ values are found to be 3.84, 3.57 and 3.22 for $x$=0.1, 0.3 and 0.4, respectively. The $\delta$ values estimated by taking the $\beta$ and $\gamma$ values from the modified Arrott plot method are 4.04, 3.45 and 3.35 for $x$=0.1, 0.3 and 0.4, respectively. Similarly, by utilizing the $\beta$ and $\gamma$ values obtained from the KF method, the $\delta$ values are estimated to be 4.06, 3.39 and 3.26 for $x$=0.1, 0.3 and 0.4, respectively. The $\delta$ values obtained using the three different strategies are very close, confirming the reliability of the critical behavior analysis. The $\delta$ value for $x = 0.1$ is remarkably larger than the value predicted by the MF theory (3) due to the large $\gamma$ value. As $x$ increases to 0.4, the $\delta$ value decreases and approaches the magnitude of the MF model. The scaling hypothesis[1] predicts that the scaled isothermals $m = M|(T-T_C)/T_C|^{-\beta}$ vs. $h = H|(T-T_C)/T_C|^{-(\beta+\gamma)}$ fall onto two different curves, one for $T > T_C$ and the other for $T < T_C$. As shown in Supplementary Figure S3 online,

the scaling hypothesis is valid for the current samples, which further confirms that the obtained critical exponent values are reliable[10,37].

**Results of electron spin resonance.** ESR has been demonstrated to be an effective tool to probe the local and microscopic magnetic states of materials. To elucidate the nature of the critical behavior around $T_C$, temperature-dependent derivative ESR spectra ($dP/dH$) were measured for $x = 0.1$ and $x = 0.3$ samples and are presented in Figure 3(a) and (b), respectively. Here, the $dP/dH$ data were normalized to the positive peak value and then shifted accordingly such that the evolution of the line shape could be easily seen. As shown in Figure 3(a) for $x=0.1$, the $dP/dH$ spectrum is asymmetrically distorted below $T_C$. When approaching $T_C$, the $dP/dH$ spectrum makes a small shift toward higher fields and becomes less asymmetrical. Once above $T_C$, however, the spectrum broadens and distorts again with increasing temperature. Synchronously, a shoulder-like feature appears on the right side of the original peak and finally evolves into a sharp peak above 235 K. Above this temperature, the two resonant peaks are clearly distinguishable in the $dP/dH$ spectra with very different peak widths. Hereafter, the original peak is denoted as $P_1$, whereas the new peak is denoted as $P_2$. As shown in Figure 3(a), $P_1$ keeps broadening as the temperature is further increased and eventually becomes invisible above 350 K, where only $P_2$ is left. As displayed in Figure 3(b), the ESR spectra for $x=0.3$ recorded with a bigger temperature interval than for $x=0.1$ show a similar evolution with temperature, even though the wide peak ($P_1$) still exists at 400 K.

The transitions in the ESR spectra discussed above can also be found in the double integrated intensity (DIN) of the original $dP/dH$ data plotted in Figure 4(a) and (b) for $x=0.1$ and $x=0.3$, respectively. For both compounds, the DIN increases remarkably as the temperature approaches $T_C$ from above. However, the inverse DIN shows an upturn at $T^{\#}$ (~ 240 K and 355 K for $x$=0.1 and 0.3, respectively), which is clearly higher than the Curie temperature, $T_C$. Above $T^{\#}$, the inverse DIN is linearly temperature dependent, suggesting a CW behavior in accordance with that observed for $1/\chi(T)$ (Figure 1).

All the $dP/dH$ data for the samples with $x$=0.1 and 0.3 can be well fitted by a sum of two Lorentzian shape functions[38], except for the data above 350 K for $x$=0.1, where a single Lorentzian function fits the data well. The fitting profiles at typical temperatures are shown in Supplementary Figure S4 online. The fitted parameters, i.e., the resonant field ($H_r$), peak-to-peak distance ($\Delta H_{PP}$) and intensity ratio of each peak to the total intensity ($S_i/\Sigma S$) for $x$=0.1 and 0.3 are presented as a function of temperature in Figure 5(a)-(f). These parameters exhibit similar trends in both compounds. As shown in Figure 5(a) and 5(b), the two ESR peaks are well separated below $T_C$ because the $H_r$ values are quite different. Above $T_C$, $H_r$ for $P_1$ increases slightly as the temperature increases and becomes nearly independent of temperature above $T^{\#}$. Meanwhile, $H_r$ for $P_2$ increases rapidly with temperature to a peak value and then decreases upon further increasing the temperature up to $T^{\#}$. Above $T^{\#}$, $H_r$ for $P_2$ maintains a constant value of 3345 Oe ($g$ ~ 2), which is attributable to PM resonance. As displayed in Figure 5(c) and (d), the $\Delta H_{PP}$ values for both peaks show a similar dependence on temperature below $T_C$. Above $T_C$, $\Delta H_{PP}$ for $P_1$ keeps increasing with temperature, whereas $\Delta H_{PP}$ for $P_2$ increases initially with temperature and then dramatically decreases at $T^{\#}$, beyond which this parameter varies little with temperature. The $S_i/\Sigma S$ values for $x$=0.1 and $x$=0.3 are shown in Figure 5(e) and (f), respectively. Below $T_C$, the $S_i/\Sigma S$ values for both peaks are comparable. $S_i/\Sigma S$ for $P_1$ is stronger than that for $P_2$ between $T_C$ and $T^{\#}$. When crossing $T^{\#}$ from below, $S_i/\Sigma S$ for $P_1$ jumps to a more dominant position at the expense of $P_2$ and then decreases gradually as the temperature increases further. Eventually, the $P_1$ peak for $x = 0.1$ disappears when $T > 350$ K (Figure 5(e)). Whereas for $x = 0.3$, the $P_1$ peak still persists even at 400 K with a reduced relative intensity (Figure 5(f)).

**Discussion**

The slightly upward deviation of $1/\chi(T)$ from the CW behavior has often been observed in normal FM materials showing stronger critical fluctuations than predicted by the MF model[37,39]. Alternative mechanisms associated with SR magnetic orders have also been proposed. In manganite oxides, this behavior has often been ascribed to the existence of SR FM clusters[18]. Phenomenologically, it is somewhat surprising because FM clusters enhance the susceptibility beyond the PM background and, in turn, lead to a reduced inverse susceptibility. In double perovskite $La_2NiMnO_6$, the AFM correlations among the neighboring FM clusters were proposed to account for the observed upward deviation in the $1/\chi(T)$ curve[21]. Because of the existence of FM clusters, the downward deviation in the $1/\chi(T)$ curve appears when a low measurement field is employed[40]. This interpretation was used to explain the upward deviation in the $1/\chi(T)$ curves observed in perovskite cobaltites $La_{1-x}Sr_xCoO_3$, for which clear evidence of SR FM clusters has been demonstrated[17]. By contrast, as shown in Figure 1(b), the insensitivity of the upward deviation in $1/\chi(T)$ to the applied magnetic field in $Cu_{1-x}NMn_{3+x}$ implies that the AFM correlations may be responsible for the deviation alone.

The peak values (~ 4020 Oe and 3630 Oe for $x=0.1$ and 0.3, respectively) of the resonant field $H_r$ for $P_2$ between $T_C$ and $T^{\#}$ are much larger than the value of PM resonance (~3350 Oe), as demonstrated in Figure 5(a) and 5(b). Such high resonant fields suggest the related magnetic coupling is AFM-type that requires a stronger magnetic field to generate the ESR resonance than the PM-type does[38,41]. However, the AFM correlations should be SR ordered because for LR AFM ordering, the strong AFM spin coupling requires a resonance field that is much stronger than the magnitude of ~ 10 kOe[42]. Above $T^{\#}$, the resonant fields for $P_1$ and $P_2$ are close to each other, thereby making them indistinguishable through bulk magnetic susceptibility measurements. As plotted in Figure 5(e) and 5(f), $P_1$ is predominantly stronger than $P_2$ in terms of intensity just above $T^{\#}$. This result implies that the magnetic coupling corresponding to $P_1$ actually provides a PM background rather than an AFM one. The upturn in $DIN^{-1}(T)$ below $T^{\#}$ (Figure 4) is a signature of the subtle reduction of DIN, which is barely visible in DIN(T). The reduction of the ESR intensity can be ascribed to the appearance of AFM couplings[43]. Therefore, the SR AFM correlations observed here can explain the upturn below $T^{\#}$ in $DIN^{-1}(T)$. The values of $T^{\#}$ below which the SR AFM correlations appear coincide well with $T^*$ (~ 250 K and 365 K for $x$ = 0.1 and 0.3, respectively), below which the magnetic susceptibility departs from the CW law, indicative of the same underlying physics for both cases. Therefore, it is the SR AFM correlations (or say, ordering) that reduce the magnetic susceptibility lower than the PM background and hence lead to the upward deviation of $1/\chi(T)$ below $T^*$, in sharp contrast to the Griffith phase case.

How do the SR AFM correlations/ordering influence the critical behavior? Ostensibly, $\gamma$ represents the divergence of the initial magnetic susceptibility upon approaching $T_C$ from above, with smaller values yielding sharper divergence[10]. In physics, $\gamma$ is in fact a measurement of the FM interaction range[8,44]. Based on the $\gamma$ values shown in Figure 2(b), the FM exchange interaction is estimated to decay as $J(r) \sim 1/r^{5.12\pm0.02}$, $\sim 1/r^{4.87\pm0.03}$ and $\sim 1/r^{4.63\pm0.02}$ for $x$=0.1, 0.3 and 0.4, respectively. Clearly, the range of FM exchange interaction increases with $x$ and approaches the anticipation of the MF model[8] (slower than ~ $1/r^{4.5}$) for $x = 0.4$. The FM exchange interaction for $x=0.1$ decays even faster than expected by the 3DH model with $J(r) \sim 1/r^5$, suggesting very strong critical fluctuations[6, 37]. The SR AFM coupled spins would force the surrounding spins to be antiferromagnetically coupled through the proximity effect[45], which prohibits the formation of FM interactions among those spins. The affection of the proximity effect on the FM magnetic couplings would decay with distance. Spatially, the SR AFM interactions associated with chemical doping should be randomly distributed. It is therefore difficult for the FM couplings to develop globally when the system is cooled toward $T_C$. As a result, the divergence of the magnetic susceptibility slows down, yielding an enhanced critical component $\gamma$. For $T_C < T < T^\#$, $H_r$ for $P_2$, which is related to the strength of the SR AFM coupling, is smaller for $x=0.3$ (Figure 5(b)) than for $x=0.1$ (Figure 5(a)). Moreover, the temperature span between $T_C$ and $T^\#$ where the SR AFM correlations appear is remarkably reduced for $x = 0.3$ (~30 K) in comparison to $x = 0.1$ (~ 60 K). This finding indicates that the SR AFM correlations are significantly weakened relative to the FM interactions with increasing $x$, leading to a reduced influence on the FM transition and hence to the suppressed critical fluctuations. Therefore, it is readily understandable that the MF-like critical behavior above $T_C$ can be restored in $x = 0.4$.

An unusually large critical component $\gamma$ has been reported in a few FM materials. In weak itinerant-electron $Zr_{1-x}Nb_xZn_2$, $\gamma$ increases from the MF magnitude to 1.33 when $x$ approaches the critical concentration $x_c = 0.083$, at which a quantum phase transition occurs[9]. However, $\beta$ always takes the value of unity as expected by the MF model. The large $\gamma$ value was also reported in perovskite $BaRuO_3$ under high pressure[6]. At ambient pressure, $BaRuO_3$ behaves as expected by the 3DH model with $\beta = 0.348$ and $\gamma = 1.41$. By applying a hydrostatic pressure, $\gamma$ increases gradually to ~ 1.8 under 0.8 GPa, whereas there is no obvious change of $\beta$. This implies enhanced critical fluctuations under pressure; however, the detailed mechanism remains elusive. In the Invar system $Fe_{100-x}Pt_x$, the critical exponent $\gamma$ was observed to be enhanced by increasing the metallurgical site disorder when the concentration was close to the stoichiometric composition $Fe_{75}Pt_{25}$; however, $\beta$ is insensitive to the disorder[8]. It was speculated that the random site disorder causes a strong broadening of the distribution of the local exchange fields owing to the competition between the strong AFM Fe-Fe exchange interaction and the dominating FM Fe-Pt and Pt-Pt interactions[8]. For manganese nitrides $Cu_{1-x}NMn_{3+x}$ with antiperovskite structure, our results clearly suggest that the SR AFM couplings appearing above $T_C$ significantly affect the critical behavior. To the best of our knowledge, the influence of SR AFM orders on the critical behavior has not been addressed in the literature. We hope the present findings can stimulate further study in this direction both experimentally and theoretically.

Figure 6 is a phase diagram for $Cu_{1-x}NMn_{3+x}$ with $x$ up to 0.5 based on our previous measurements of the bulk magnetic susceptibility and temperature-dependent X-ray diffraction[35]. When $x$ increases, the structural transition temperature $T_S$ decreases and finally disappears above $x$=0.4; however, $T_C$ shifts toward higher temperatures. Similar trends were reported by Takenaka et al in nitrogen-deficient $CuN_{1-\delta}Mn_3$[34], where the nitrogen deficiency causes a reduction of $T_S$ and an increase of $T_C$. Based on the results discussed above, a temperature zone of SR AFM ordering against the PM background can be added to the phase diagram, which isolates the low-temperature FM ordered state from the high-temperature PM state.

The cubic antiperovskite lattice is composed of corner-sharing $Mn_6N$ octahedra (see Supplementary Figure S1(a) online) and contains 3D geometrical frustration in terms of AFM interactions[46]. In most cases, the AFM structure in cubic antiperovskite manganese nitrides takes triangular configurations[47,48], where Mn moments on the (111) plane point 120° away from each other (see Supplementary Figure S1(b) online as an example). Complex AFM structures, e.g., a square configuration along with a FM component in a tetragonal crystal lattice or the existence of two sets of collinear AFM configurations, were observed in a very small number of compounds[47,48]. For the current case, the complex AFM structures may be excluded because only a single AFM signal was observed in addition to the PM background in the ESR spectra. Moreover, the AFM structures reported thus far in chemically doped $CuNMn_3$ compatible with a cubic crystal structure are all in triangular configurations[48]. In this context, the SR AFM ordering observed in $Cu_{1-x}NMn_{3+x}$ probably takes a triangular configuration. Although it is not favored by the geometrical frustration, the SR AFM ordering can still be expected if frustration is locally relieved by the lattice deformation due to the random substitution of Cu sites with Mn atoms. The local structural distortion upon doping has been verified to play a very important role in determining the physical properties and functionalities of antiperovskite manganese nitrides[23,24,49-51]. As reported in Ref. 23, strong frustration is suggested to be associated with lattice contraction in the PM state. According to our previous work[35], the lattice constant keeps decreasing with increasing $x$, suggestive of strengthening geometrical frustration globally. Separately, the corner-site Mn atoms that replace the Cu sites are suggested to be magnetically coupled with the face-center-site Mn atoms by isotropic exchange interactions, which may disturb the AFM interactions within the (111) plane[52]. Because of the strengthening geometrical frustration and the disturbance from the corner-site Mn atoms, the local AFM ordering is unable to survive at higher doping levels, and the MF-type critical behavior is finally recovered at $x$=0.4. For further study, local structural probes, such as neutron or X-ray pair distribution function and X-ray absorption fine structure spectra, are desirable. As in the situation of strongly correlated perovskite oxides, our result along with previous reports on local structure distortion suggests that the intrinsic magnetic, electronic and structural inhomogeneity on the microscopic scale is crucial for comprehending the bulk physical properties of antiperovskite manganese nitrides.

In summary, we studied the magnetic properties of $Cu_{1-x}NMn_{3+x}$ ($x$=0.1, 0.2, 0.3 and 0.4) using magnetic susceptibility and ESR measurements. We found that the inverse magnetic susceptibility $1/\chi(T)$ exhibits an upward deviation from the CW behavior above $T_C$, which was attributable to the existence of SR AFM correlations, as demonstrated by the ESR result. For all compositions, the critical behavior below $T_C$ was well described by the MF model. However, the critical fluctuations in compounds with small $x$ values were much stronger than predicted by the MF model. As $x$ increases, the critical fluctuations were suppressed gradually, and the MF-like behavior was recovered for $x$=0.4. SR AFM correlations were proposed as the cause of the faster decay of the FM interaction length than that expected by the MF model, leading to the unusual critical behavior above $T_C$.

**Methods**

Polycrystalline samples of $Cu_{1-x}NMn_{3+x}$ with $x$=0.1, 0.2, 0.3 and 0.4 were prepared via a direct solid-state reaction. Powders of Cu (4 N), Mn (4 N) and homemade $Mn_2N$ were mixed in the desired ratios, pressed into pellets, sealed in evacuated quartz tubes (~$10^{-6}$ torr) and sintered at 750 ºC for 3 days, followed by annealing at 800 ºC for 5 days. After quenching the tubes to room temperature, the products were ground carefully, pressed into pellets, sealed in evacuated tubes and annealed again at 800 ºC for an additional 8 days. The Mn$_2$N powders were prepared using the following sequence. First, the Mn (4N) powders were wrapped with Ta foil and placed into a high-pressure pipe furnace made of nickel-based stainless steel. Next, the pipe furnace was repeatedly cleaned using high-purity nitrogen gas (5N) before being sealed with high-purity nitrogen gas inside. Then the furnace was heated to 750 ºC and held for 60 hours while the inside-tube gas pressure was maintained as 5 MPa. Finally, high-quality Mn$_2$N powders were obtained. From X-ray diffraction, all the samples exhibited the antiperovskite structure (group symmetry, $Pm\bar{3}m$) without a detectable second phase. For the $x$=0.1 and 0.3 samples, the magnetic measurements were performed on a Quantum Design superconducting quantum interference device (SQUID). For the sample with $x$=0.4, the measurements were carried out on a vibrating sample magnetometer with an oven attached on a Quantum Design physical property measurement system (PPMS). The isothermal magnetizations ($M$-$H$ curves) were measured after the samples were cooled down to each measurement temperature from well above $T_C$. The electron spin resonance (ESR) spectra were recorded using an X-band Bruker EMX plus 10/12 cw spectrometer operating at 9.4 GHz. ESR detects the power $P$ absorbed by a sample from a transverse microwave magnetic field. The signal-to-noise ratio of the spectra was improved by recording the first derivative of $P$ ($dP/dH$) using a lock-in technique.

**Figure captions**

**Figure 1 │ Inverse magnetic susceptibility 1/χ(T) for $Cu_{1-x}NMn_{3+x}$ (x=0.1, 0.2, 0.3 and 0.4).** (a) 1/χ(T) at 0.1 kOe for all samples. (b) 1/χ(T) measured at 0.1 kOe, 1 kOe, 3 kOe and 10 kOe for the x=0.2 sample. The onset temperatures ($T^*$) where 1/χ(T) deviates from the linear temperature dependence (solid lines) are indicated by the arrows. The ferromagnetic transition for each composition at $T_C$ is marked as a solid diamond in (a).

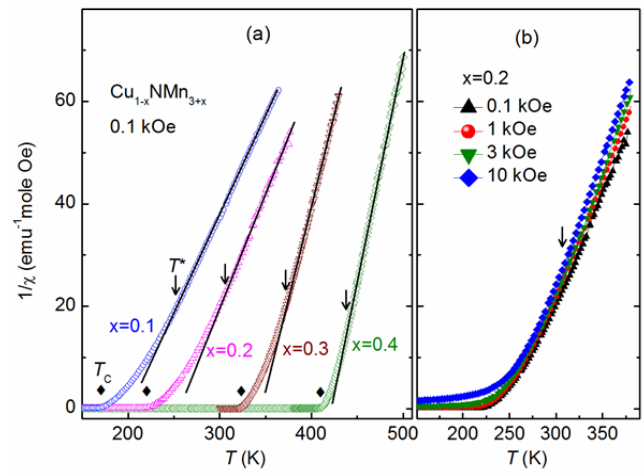

**Figure 2 | Critical behavior analysis for Cu$_{1-x}$NMn$_{3+x}$ (x=0.1, 0.3 and 0.4).** (a) The spontaneous magnetization $M_s$ (left) and inverse initial magnetic susceptibility $1/\chi_0$ (right) vs. ($T-T_C$) along with the fit (solid lines) to the power laws (see the text for details). The obtained critical exponents ($\beta$, $\gamma$) are shown. (b) The Kouvel-Fisher plot of $M_s$ (left) and $1/\chi_0$ (right) as a function of $T-T_C$. The solid lines represent the linear fitting of the data. The obtained critical exponents ($\beta$, $\gamma$) are shown. (c) Critical isotherm $M(H)$ on log-log scale at the temperature closest to $T_C$ for Cu$_{1-x}$NMn$_{3+x}$ (x=0.1, 0.3 and 0.4). The solid lines are linear fits, and the obtained critical exponent $\delta$ is shown for each sample.

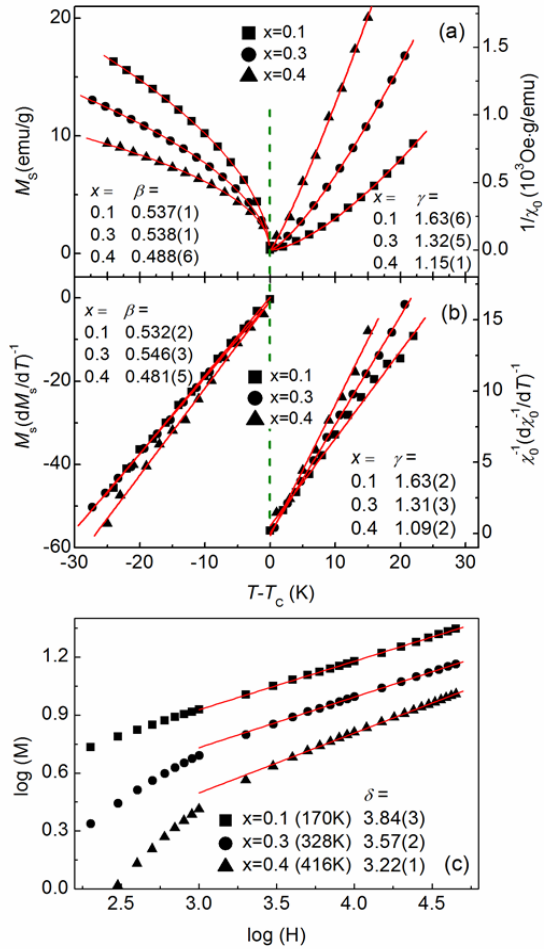

**Figure 3 | ESR spectra, *dP/dH*, as a function of temperature for $Cu_{1-x}NMn_{3+x}$.** (a) *dP/dH* for *x*=0.1. (b) *dP/dH* for *x*= 0.3. Note that the spectra plotted here were normalized and shifted (see the text for details). The dashed lines indicate the evolutions of resonant peaks ($P_1$ and $P_2$) with temperature above $T_C$.

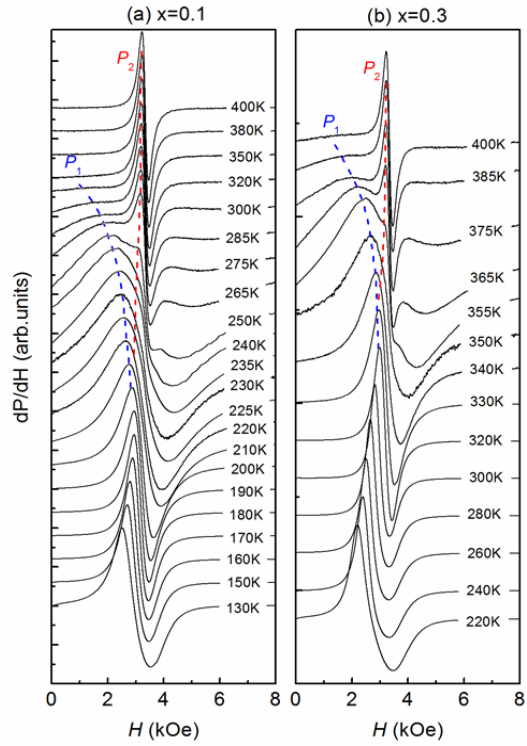

**Figure 4 │ Double integrated intensity (DIN) of the original ESR spectra for $Cu_{1-x}NMn_{3+x}$.** (a) for $x=0.1$ and (b) for $x=0.3$. The inverse DIN is also shown in each panel. The solid line (red) on $DIN^{-1}(T)$ curve indicates a linear fit. The temperatures ($T^{\#}$s) at which $DIN^{-1}$ departs from the linear dependence on temperature and the Curie temperature $T_C$ are indicated by the arrows.

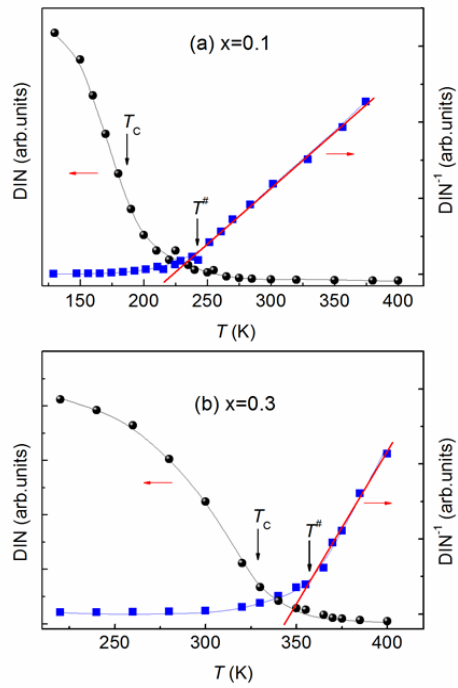

**Figure 5 | Temperature dependence of the fit parameters, i.e., the resonant field ($H_r$), peak-to-peak distance ($\Delta H_{PP}$) and intensity ratio of each peak to the total intensity ($S_i/\Sigma S$) of the ESR spectra for $Cu_{1-x}NMn_{3+x}$.** (a) $H_r$ for $x=0.1$. (b) $H_r$ for $x=0.3$. (c) $\Delta H_{PP}$ for $x=0.1$. (d) $\Delta H_{PP}$ for $x=0.3$. (e) $S_i/\Sigma S$ for $x=0.1$. (f) $S_i/\Sigma S$ for $x=0.3$. The temperatures $T_C$ and $T^\#$ are indicated by the vertical dotted lines. The solid lines are guides to the eyes. $P_1$ and $P_2$ refer to the resonant peaks above $T_C$.

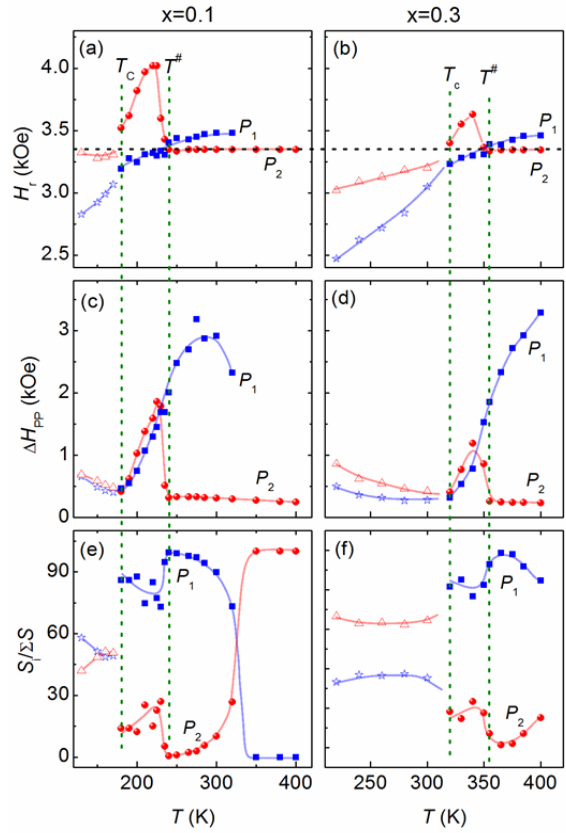

**Figure 6 | Phase diagram for Cu$_{1-x}$NMn$_{3+x}$ with $0 \leq x \leq 0.5$.** $T_C$ represents the ferromagnetic (FM) Curie temperature, and $T_S$ represents the tetragonal-cubic structural transition temperature. $T^*$ denotes the temperature below which $1/\chi(T)$ deviates from the Curie-Weiss law, and $T^\#$ is the onset temperature below which short-range antiferromagnetic ordering (SR AFM) occurs and coexists with the paramagnetic (PM) matrix.

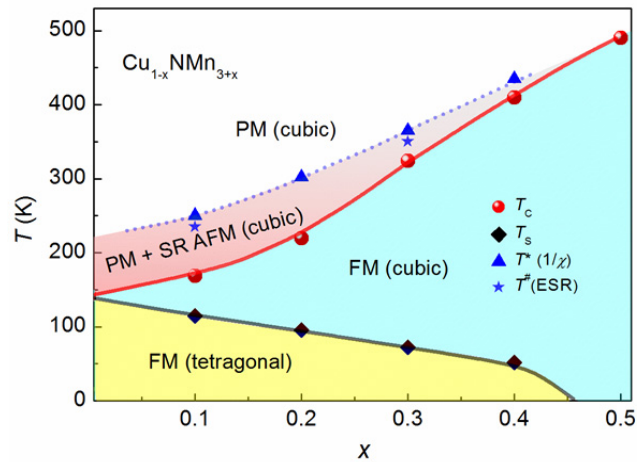

## Acknowledgments

This work was supported by the National Key Basic Research under Contract No. 2011CBA00111; the National Natural Science Foundation of China under Contract Nos. 51322105, 11174295, 51301167, 51171177 and 91222109; the Joint Funds of the National Natural Science Foundation of China and the Chinese Academy of Sciences' Large-Scale Scientific Facility (Grant No. U1232138); and the Foundation of Hefei Center for Physical Science and Technology under Contract No. 2012FXCX007. The authors thank Dr. Chen Sun for her assistance in editing the manuscript.

## Author contributions

J. C. L prepared the samples and performed the characterization. J. C. L., D. P. C., C. Y., S. L. and B. S. W. performed the magnetic susceptibility experiments. P. T and L. Z. carried out the critical behavior analysis and data interpretation. W. T., Y. M. Z. and J. Y. assisted with the ESR measurement and contributed to the data analysis and interpretation. P. T and Y. P. S. designed the experiments and guided the work. P. T. wrote the manuscript with help from the co-authors. All the authors discussed the results and reviewed the manuscript.

**Additional information**

Supplementary Information statement.

Competing financial interests: the authors declare no competing financial interests.